\documentclass[nofootinbib,aps,preprint,prd]{revtex4-1}

\usepackage{graphicx}
\usepackage{amssymb}
\usepackage{bbm}
\usepackage{color}
\usepackage{ulem}
\usepackage{slashed}

\newcommand{\ben}{\begin{enumerate}}
\newcommand{\een}{\end{enumerate}}
\newcommand{\bit}{\begin{itemize}}
\newcommand{\eit}{\end{itemize}}

\newcommand{\beqa}{\begin{eqnarray}}
\newcommand{\eeqa}{\end{eqnarray}}
\newcommand{\beq}{\begin{equation}}
\newcommand{\eeq}{\end{equation}}
\newcommand{\bay}{\begin{array}}
\newcommand{\eay}{\end{array}}

\newcommand{\ra}{\rangle}
\newcommand{\la}{\langle}
\newcommand{\vev}[1]{\la{#1}\ra}

\newcommand{\nn}{\nonumber}

\def\gsim{\ \rlap{\raise 3pt \hbox{$>$}}{\lower 3pt \hbox{$\sim$}}\ }
\def\lsim{\ \rlap{\raise 3pt \hbox{$<$}}{\lower 3pt \hbox{$\sim$}}\ }

\def\ord{{\cal O}}
\def\lt{\left}
\def\rt{\right}

\newcommand{\Tr}{{{\rm Tr}}}

\def\eg{{\it e.g.}}
\def\ie{{\it i.e., }}
\def\lag{{\cal L}}
\def\CC{{\cal C}}
\def\HH{{\cal H}}
\def\EE{{\cal E}}

\def\bracket#1#2 {\mathinner{\langle{#1}|{#2}\rangle}}

\def\id{\mathbbm{1}}

\def\OMIT#1{{}}

\begin{document}

\title{\ \\ Spontaneous CP Violation and Light Particles\\ in The Littlest Higgs\\\ }


\author{Ze'ev Surujon}
\email[]{zevik@ucsd.edu}
\affiliation{Department of Physics, University of California at San Diego, La Jolla, CA 92093}

\author{Patipan Uttayarat}
\email[]{puttayarat@physics.ucsd.edu}
\affiliation{Department of Physics, University of California at San Diego, La Jolla, CA 92093}

\preprint{UCSD PTH 10-02}

\begin{abstract}
Little Higgs models often feature spontaneously broken extra global symmetries,
which must also be explicitly broken in order to avoid massless Goldstone modes
in the spectrum.
We show that a possible conflict with collective symmetry breaking
then implies light modes coupled to the Higgs boson, leading to interesting
phenomenology.
Moreover, spontaneous CP violation is quite generic in such cases,
as the explicit breaking may be used to stabilize physical CP odd phases
in the vacuum.
We demonstrate this in an SU$(2)\times$SU$(2)\times$U$(1)$
variant of the Littlest Higgs, as well as in an original SU(6)/SO(6) model.
We show that even a very small explicit breaking may lead to large phases,
resulting in new sources of CP violation in this class of models.
\end{abstract}
\maketitle

\newpage

\section{Introduction}

Despite its impressive experimental success, the Standard Model (SM) is known to have
several theoretical puzzles.
One of these, the ``hierarchy problem'', is the apparent fine tuning associated with the
electroweak scale.
This paradigm has led to numerous hypotheses, such as supersymmetry, technicolor,
extra-dimensions, and more.
In order to eliminate the hierarchy problem, models based on these hypotheses often introduce
new physics at the TeV scale.
Unfortunately, such low scale new physics seem to generally spoil
the success of the SM by introducing low energy effects which are
tightly constrained by experimental data.
%
%

The tension between the need to solve the hierarchy problem and the above experimental
constraints is known as the ``little hierarchy problem''.
It may be solved by using the Little Higgs framework~\cite{LH},
where physics beyond the SM appears only at $\Lambda\sim 10$~TeV
instead of the generically expected $1$~TeV.
The SM Higgs field remains naturally light by serving as a pseudo Goldstone boson
of multiple approximate global symmetries.
Explicit breaking of this set of symmetries is ``collective'', \ie
apparent only in the presence of at least two terms in the Lagrangian.
This ensures that the only one-loop diagrams contributing to the Higgs mass
are logarithmically divergent at most, thereby allowing for a cutoff at
$\Lambda\sim (4\pi)^2v$ instead of the generic $\Lambda\sim 4\pi v$.
%


In Little Higgs models, the electroweak gauge group is extended to a partially gauged
global symmetry.
The gauged generators are broken spontaneously to the electroweak gauge group.
Some of the global generators are broken spontaneously too, but in a realistic model
they must be also broken explicitly in order to avoid exact Goldstone bosons.
Then, one has to make sure that the set of global symmetries which protect the Higgs is not
broken non-collectively.
Such non-collective breaking would destabilize the electroweak scale.

In this paper we discuss cases, such as the SU$(2)^2\times$U$(1)$ Littlest Higgs
variant~\cite{PPP}, where there is a tension between lifting the mass of the
pseudo-Goldstone bosons and retaining collective symmetry breaking.
A consequence of this is the presence of light
particles with direct couplings to the SM Higgs, leading to
interesting phenomenology.
For example, there is a range of parameters for which
a new decay channel for the Higgs opens up.

Another possible consequence is the appearance of spontaneous CP violation,
\ie physical phases in the VEV.
Such phases are rotated by field redefinitions.
The generators of these transformations must obey some conditions if the vacuum indeed breaks CP
invariance~\cite{SCPV-cond}.
In particular, in order for a phase to be physical, the related generator must be
both explicitly and spontaneously broken.
In case there is a conflict between this requirement and that of collective
symmetry breaking,
one may expect that the effect of spontaneous CP violation is suppressed - from
the same reason that the related pseudo-Goldstone bosons are light.
However, as we will show, the CP violating phase may be $\ord(1)$, even in the limit of
small explicit breaking.
We begin by reviewing the Littlest Higgs model and its SU$(2)^2\times$U$(1)$ variant,
showing that it includes an exact Goldstone due to a spontaneously broken global U$(1)$
which would be gauged in the original SU$(2)^2\times$U$(1)^2$ version of the littlest Higgs.
%
%
We then show that lifting the exact Goldstone requires spoiling collective symmetry breaking, hence leading to a suppression of its mass.
%
%
Once collective symmetry breaking is spoiled, even by a small parameter,
it becomes possible for the vacuum to align with an $\ord(1)$ CP-odd phase.
We discuss how such CP violation arises in the low energy
limit.   In the SU$(2)^2\times$U$(1)$ model it turns out to be suppressed,
but we argue that
this is a peculiarity of the minimal nature of the SU$(5)$ structure, rather than
a generic feature in Little Higgs models.
In order to support this statement, we construct an original SU(6)/SO(6) model
which accommodates an $\ord(1)$ physical phase in the low energy limit.
%

%

\section{Saving the SU$(2)\times$SU$(2)\times$U$(1)$ Model}
\label{sec:LLH}

Here we will discuss the SU$(2)\times$SU$(2)\times$U$(1)$, and how
to make its Goldstone boson massive without destabilizing the electroweak scale.
But before that, let us briefly review the original Littlest Higgs.

\subsection{The Littlest Higgs}
A very elegant implementation of the Little Higgs idea is the Littlest 
Higgs~\cite{LLH},
%
%
whose lagrangian is described as an approximate SU$(5)/$SO$(5)$ effective field theory.
The vacuum manifold SU$(5)/$SO$(5)$ may be parametrized as $\Sigma_0=UU^T$, where $U$ is
a broken SU$(5)$ transformation.
The global SU$(5)$ is explicitly broken by gauging
an $[{\rm SU}(2)\times {\rm U}(1)]^2$ subgroup, where the gauged generators are embedded in SU(5) as
\beqa
   T_1^i &=& \pmatrix{\sigma^i/2 & & \cr & 0_{1\times1} & \cr & & {0}_{2\times 2}}, \quad
   Y_1={\rm diag}(3,3,-2,-2,-2)/10; \nonumber\\
   T_2^i &=&  \pmatrix{{ 0}_{2\times 2} & & \cr & 0_{1\times1} & \cr & & -\sigma^{a*}/2}, \quad
   Y_2={\rm diag}(2,2,2,-3,-3)/10.
\eeqa
Once this explicit SU$(5)$ breaking is included, the degeneracy is partially lifted, as
a minimum energy vacuum appears at
\beq
	\label{eq:VEV-Sigma}
   \Sigma_0=\pmatrix{ & & e^{i\delta}V\cr & e^{-4i\delta} & \cr e^{i\delta}V^T & &}.
\eeq 
Here, $V$ is a $2\times2$ special unitary matrix and $\delta$ is a real parameter.
%
%
Gauging the $[{\rm SU}(2)\times {\rm U}(1)]^2$ subgroup
breaks explicitly all the SU$(5)$ generators which are not gauged.
The vacuum breaks the $\lt[{\rm SU}(2)\times{\rm U}(1)\rt]^2$ gauge group
to the electroweak group, SU$(2)_L\times$U$(1)_Y$.
One can then use the spontaneously broken generators to rotate the vacuum into the form
\beq
   \label{eq:VEV-LLH}
   \Sigma_0=\pmatrix{ & & \id\cr & 1 &\cr \id & & }.
\eeq
By doing so, we have chosen a basis in which the electroweak gauge group
is given by the diagonal subgroup of the full Little Higgs gauge group.
%
%

We follow the common formalism for chiral lagrangians~\cite{CCWZ} and arrange the Goldstone bosons in
a matrix,
\beq
   \Sigma=e^{i\Pi/f}\Sigma_0 e^{i\Pi^T/f}, \qquad \Pi=\Pi^a X^a,
   \label{eq:sigma}
\eeq
where $X^a$ are the 14 broken  generators of SU(5).
We can always choose a basis where 
\beq
   X^a\Sigma_0=\Sigma_0 X^{aT} \label{eq:broken}.
\eeq
In this basis, Eq.(\ref{eq:sigma}) simplifies to
\beq
   \Sigma=e^{2i\Pi/f}\Sigma_0.
\eeq
Four of the above fourteen degrees of freedom become the longitudinal
components of the $W^{'\pm}$, $Z'$ and $\gamma'$, which correspond
to the spontaneously
broken gauged generators.  The remaining ten pseudo-Goldstone bosons
can be classified according to their SM quantum numbers as one complex
doublet $H$, which we identify with the SM
Higgs, and one complex triplet, $\phi$, which carries one unit of
hypercharge.  These pseudo-Goldstone bosons are parametrized as
follows:
\beq
   \Pi=\pmatrix{ {\rm eaten}& H/\sqrt2 & \phi\cr
          H^\dagger/\sqrt2 & {\rm eaten}& H^T/\sqrt2 \cr
          \phi^\dagger & H^\ast/\sqrt2 & {\rm eaten}\cr}.
\eeq

From the transformation law $\Sigma\to U\Sigma U^T$, it follows
%
%
that the Higgs transforms nonlinearly under SU$(3)_1$ and
SU$(3)_2$, which act on the $(123)$ and $(345)$ blocks, respectively.
Note that the SU$(2)_1\times $U$(1)_1$ gauge interactions break
SU$(3)_1$ and conserve SU$(3)_2$, whereas SU$(2)_2\times $U$(1)_2$
gauge interactions conserve SU$(3)_1$ and break SU$(3)_2$.
However, the two (overlapping) groups SU$(3)_1$ and SU$(3)_2$ are fully broken
only when both sets of gauge couplings are turned on, namely, they are collectively broken.
Therefore, any diagram which contributes
to the Higgs mass must involve both `1' and `2' gauge interactions.
However, the only one-loop diagrams contributing to the Higgs mass
involve two gauge boson propagators,
leading to only a logarithmic dependence:
$\delta m_H^2\sim\lt(\frac{gf}{4\pi}\rt)^2\log(\Lambda/f)$. 

In order to maintain collective symmetry breaking also in the top quark sector,
we introduce a new vector-like quark pair $(t'_L,t'_R)$ which is SU$(2)_L$ singlet,
and we define $\chi^i_L=(i\sigma^2 Q_L,t'_L)$, where $i=1,2,3$ and $Q_L$ is the SM third generation quark
doublet.
The top quark sector is taken to be
\beq
   \label{eq:yukawa}
   \lag=\lambda f\overline{\chi}_{Li} \Omega^i t_R + \lambda' f  \bar t'_L t'_R + {\rm c.c.},
\eeq
where
\beq
   \Omega^i = \epsilon^{ijk}\epsilon^{xy}\Sigma_{jx}\Sigma_{ky}.
   \label{eq:Omega}
\eeq
Here,  $i,j,k$ run over $1,2,3$ and  $x,y$  over $4,5$.
The first term is invariant under SU$(3)_1$, but breaks SU$(3)_2$,
whereas the second term breaks SU$(3)_1$ and preserves SU$(3)_2$.
That this is the case can be seen by taking $\chi_L^i$ to be an SU$(3)_1$ triplet.
Diagrams which contribute to the Higgs mass must involve both couplings,
and are only logarithmically divergent at one-loop.
%

\subsection{The Hypercharge Model}

The Littlest Higgs model suffers from large corrections to
electroweak precision observables, mainly due to the heavy gauge boson
related to $U(1)'$.
One solution to this problem is to impose T-parity~\cite{LHT}, under which SM fields are even and
new heavy fields are odd.   This removes all the single heavy field
exchange diagrams, effectively pushing many dangerous contributions to
the electroweak precision observables to the loop level.
In the same time, T-parity provides a WIMP dark matter which naturally gives the correct
thermal relic abundance.  
Nevertheless, the multitude of new fields makes it potentially vulnerable to
flavor problems~\cite{XCPV}.
Moreover, it becomes
difficult to find a simple UV completion to match it onto~\cite{UV-comp}.

Another solution was to gauge only one of the U$(1)$ generators~\cite{PPP}.
It is this solution that we are considering here, although
our lessons for model building and spontaneous CP violation are
rather generic, and we expect them to hold whether or not T-parity is imposed.
Let us define the following two combinations of U$(1)$ generators:
\beqa
   Y &=& \frac{Y_1+Y_2}{2}=\frac{1}{2}{\rm diag}(1,1,0,-1,-1),\nonumber\\ && \nonumber\\
   Y' &=& \frac{Y_1-Y_2}{2}=\frac{1}{10}{\rm diag}(1,1,-4,1,1).
\eeqa
In this model which we denote as the {\it hypercharge model}, only the the
SM hypercharge $Y$ is gauged while $Y'$ generates a global symmetry, which we denote
as U$(1)'$.  
The pseudo-Goldstone bosons matrix now becomes (in terms of the uneaten fields)
\beq
   \Pi=\pmatrix{ \eta/\sqrt{20}\, \id& H/\sqrt2 & \phi\cr
          H^\dagger/\sqrt2 & -2\eta/\sqrt{5}& H^T/\sqrt2 \cr
          \phi^\dagger & H^\ast/\sqrt2 & \eta/\sqrt{20}\, \id\cr}.
\eeq
%

While gauging U$(1)_Y$ alone eliminates the troublesome heavy gauge boson,
it spoils collective symmetry breaking, since unlike U$(1)_1$ or U$(1)_2$ gauge
interactions which conserve one SU$(3)$ each,
%
%
the hypercharge gauge interaction breaks explicitly both SU$(3)_1$ and SU$(3)_2$ via a single
term in the Lagrangian.
As was shown in~\cite{PPP}, this effect is suppressed by the smallness of the
hypercharge coupling $g'$, and we will not discuss it further.\footnote{
Radiative corrections break SU$(3)_1$ already in the original
$\lt[{\rm SU}(2)\times{\rm U}(1)\rt]^2$ model~\cite{GKU}.
Here we will consider only tree-level breaking.}

Another, more acute problem of the hypercharge model is that it introduces a new massless
Goldstone boson $\eta$, which corresponds to the spontaneously broken U$(1)'$.
Note that so far, U$(1)'$ is an exact symmetry which is only broken spontaneously.
In the Littlest Higgs, this Goldstone boson is eaten by the corresponding gauge boson, which
is absent in the hypercharge model.
Therefore, in order for this model to be phenomenologically viable, the new Goldstone must
acquire mass, requiring explicit breaking of U$(1)'$.  
This has been recognized previously, but without providing explicit realization.
For example, in~\cite{LH-pseudoscalars}, the phenomenology
of $\eta$ was studied, {\it assuming} a range of masses up to $m_\eta\sim v$.
Below we show that any operator that gives mass to $\eta$ is bound to introduce
further non-collective symmetry breaking, thus constraining $m_\eta$ to be
roughly below the SM Higgs mass.
The assumption on $m_\eta$ in~\cite{LH-pseudoscalars} is therefore consistent with
our results.

\subsection{A Realistic Hypercharge Model}
\label{sec:PPP}

Before proving that collective symmetry breaking must be spoiled
by any term that breaks U$(1)'$, let us state a generic condition
any explicitly broken generator has to satisfy
in order not to spoil collective symmetry breaking, namely, in order not to break the full set of
symmetries which protect the Higgs by a single term in the lagrangian.
In order to do that, denote the collection of groups under which the Higgs transforms
non-homogeneously by $\{\CC_i\}$.
Each of these groups should be {\it minimal} in the sense that it does not contain a subgroup
which protects the Higgs.
The $\CC_i$ may be disjoint (as in the Minimal Moose~\cite{MM} or in
the Simplest Little Higgs~\cite{SLH}) or overlapping (as in the Littlest Higgs, where
we have $\CC_1=\text{SU}(3)_1$ and $\CC_2=\text{SU}(3)_2$).

Consider a generator $X$.
First note that if $X$ is a linear combination of gauged generators and a generator of $\CC_i$ for a particular
$i$, then breaking $X$ explicitly requires breaking $\CC_i$ explicitly too (since gauge invariance must be an
exact symmetry).
If this is true for all $i$, then any term in the lagrangian which breaks $X$ explicitly would inevitably spoil
collective symmetry breaking.
We thus arrive at the following condition:\\ \\
{\it In order that a generator can be broken explicitly without spoiling collective symmetry
breaking, it must not be expressible as any kind of the linear combinations above.}\\ \\
Failing to satisfy this condition would lead to non-collective breaking of the set $\{\CC_i\}$,
which may be allowed provided that the breaking is small enough, such that it does not destabilize
the weak scale.

Applying the condition above to $Y'$, we see that
the generator $Y'$ cannot be broken without spoiling collective symmetry since
it can be expressed as:
\beq
   5Y'=-Y+2\sqrt3 T_{\text{SU}(3)_1}^8= Y+2\sqrt3 T_{\text{SU}(3)_2}^8.
\eeq
Thus any term which breaks U$(1)'$ and is allowed by gauge invariance must break
both SU$(3)_1$ and SU$(3)_2$.
A spurion which qualifies is $s=(0,0,1,0,0)^T$, transforming (formally) in the fundamental
of SU$(5)$.
Its symmetry breaking pattern is SU$(5)\to$SU$(4)$ which acts on the $(3,3)$ minor.
The $9$ broken generators include $Y'$ and generators which are also
broken by the gauging.
In particular, any function of $\Sigma_{33}=s^\dagger\Sigma s$
would break $Y'$ while maintaining gauge invariance.
For example, consider
\beq
   \delta\lag=\varepsilon f^4\Sigma_{33}+{\rm c.c.},
   \label{eq:sigma33}
\eeq
where $\varepsilon$ is dimensionless.
Expanding $\Sigma$, we obtain
\beq
   \delta\lag=4\varepsilon f^2\lt[\frac{4}{5}\eta^2+H^\dagger H+\ldots\rt],
\eeq
where we took $\varepsilon$ to be real, such that no extra explicit CP violation is implied.
As expected, $m_H$ gets a tree-level contribution, since $\delta\lag$ breaks
explicitly both SU$(3)_1$ and SU$(3)_2$.
In order not to destabilize the electroweak scale, we require
$\varepsilon\sim (1/4\pi)^2$.

It follows that mass of $\eta$ can be as large as the Higgs mass, but
it seems equally reasonable (or equally unreasonable)
to have a much lighter $\eta$.

\begin{figure}
	\centering
	\includegraphics[width=.45\textwidth]{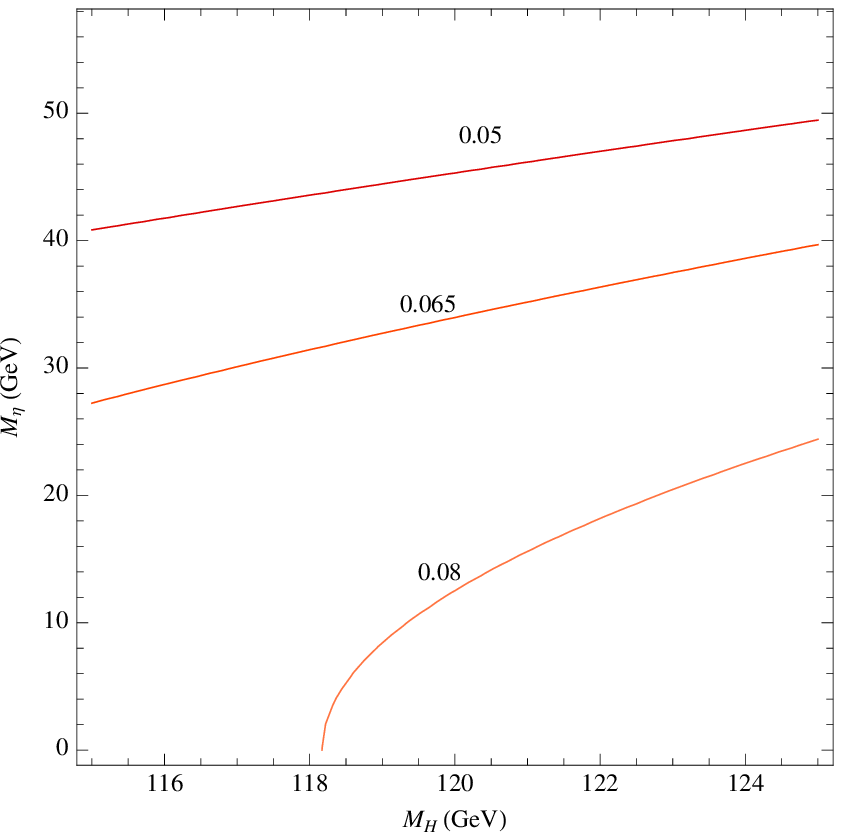}
	\includegraphics[width=.45\textwidth]{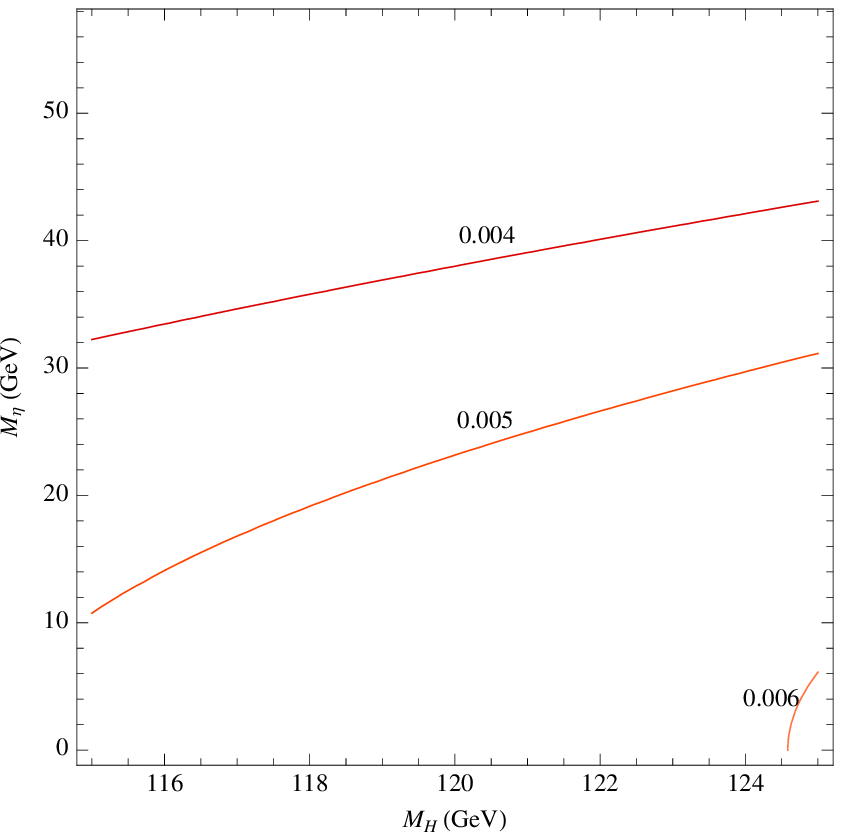}
	\caption{\footnotesize Contours of the branching ratio $BR(h\to\eta\eta)$
	in the $m_\eta-m_H$ plane, for $f=1.2$~TeV (left) and $f=2.4$~TeV
	(right).  While the dominant mode in this mass range is still $b\bar b$,
	the new mode $\eta\eta$ becomes significant throughout the parameter
	space.	\label{fig:ratioSU5}}
\end{figure}

A light $\eta$ which couples directly to the Higgs [via both a renormalizable term
$\sim\varepsilon\eta^2 H^\dagger H$ and derivative couplings
such as $\sim\frac{1}{f^2}(\eta\partial^\mu\eta)(H^\dagger\partial_\mu H)$]
would open a new decay channel $h\to\eta\eta$ for the SM Higgs
(see fig.~\ref{fig:ratioSU5}).
%
%
%
%
Due to its sizable couplings, $\eta$ would decay promptly at the collider, but
depending on its dominant decay modes, it could lead to unusual signatures.
For example, $\eta$ can decay into a pair of light particles: $e^+e^-$,
or $b\bar b$, as studied in~\cite{LH-pseudoscalars}.
In this case, the Higgs can decay into two pairs of boosted objects, such as
$H\to \eta\eta\to (jj)+(\ell^+\ell^-)$, where the objects in the parentheses
are collimated due to the large boost factor $\gamma\sim m_H/2m_\eta$.
Another possibility is that the singlet $\eta$ decays mainly via off-shell top quarks.
Then, the Higgs will decay into two pairs of boosted tops.
%
The viability of such unusual 
Higgs phenomenology deserves careful
study, which we leave for future work.

%

%

%

The explicit breaking could as well be introduced in the Yukawa term, in which case the bound on
$\varepsilon$ comes from one-loop.  
For example, consider the following Yukawa term:
\beq
        \lag_Y = f\bar \chi_{Li}\Omega^i\lt(\lambda +\varepsilon\Sigma_{33}\rt)
        t_R+{\rm c.c.},
\eeq
inducing a one-loop contribution of the form
\beq
   \kappa\varepsilon^2 f^2 \lt(\frac{\Lambda}{4\pi}\rt)^2\Omega^\dagger\Omega
   \lt(\lambda^2+2\lambda\varepsilon{\cal R}e\Sigma_{33}+\varepsilon^2\lt|\Sigma_{33}\rt|^2\rt),
\eeq
where $\kappa$ is an order one number depending on physics near the UV cutoff,
and $\varepsilon$ is real valued.
This gives rise to
\beq
   m_{\eta}\sim m_H\sim \varepsilon \lambda f.
\eeq
Therefore $\varepsilon$ may be as large as $\sim 1/4\pi$.

In the next section we discuss how such explicit breaking of U$(1)'$,
when properly introduced, may give rise to spontaneous CP violation, by 
stabilizing the phase $\delta$ in Eq.(\ref{eq:VEV-LLH}),
such that the VEV assumes the form
\beq 
   \Sigma_0 = e^{i\varphi}
   \pmatrix{ & & e^{i\delta}\id\cr & e^{-4i\delta} & \cr  e^{i\delta}\id & &} .
   \label{eq:VEV-PPP}
\eeq
%

\section{Spontaneous CP Violation in the Hypercharge Model}

\label{sec:SCPV}

\subsection{Spontaneous CP Violation from Breaking U$(1)'$}

We saw that the Hypercharge model has an explicitly broken U$(1)'$.
One might expect that once this U$(1)'$ is broken (for example, by
$\Sigma_{33}$ insertions, like in the previous section),
the related pseudo-Goldstone $\eta$ acquires a VEV that breaks CP spontaneously.
It turns out that there must be at least two terms which break U$(1)'$ explicitly,
in order for spontaneous CP violation to occur~\cite{SCPV-cond}.
It can be shown generically that as long as a U$(1)$ is broken by one single term
involving a single field,
the U$(1)$-related phase in the VEV can be removed, by using a particular U$(1)$ transformation.
Because of the explicit breaking, the coupling flips its sign, but the phase is removed.\footnote{
%
We thank H.~Haber for pointing out such a possibility.  See more examples
in~\cite{SCPV-cond}.}
Once two explicit breaking terms are introduced, the phase gets generically
stabilized at a non-zero value.
Of course, more terms would be induced by loops, but the resulting phase
would be also loop suppressed.

It is interesting to notice that the CP-odd phase may be $\ord(1)$, even in the limit
$\varepsilon\to 0$, where the explicit breaking vanishes.
The reason for this non-analytical behavior of the phase as function of $\varepsilon$
is that however small $\varepsilon$ may be, it is the leading effect in lifting the degeneracy
associated with the Goldstone direction.
Nevertheless, any 
physical consequence of effect related
to the phase is associated with some momentum scale $p$ (for example, the mass of
a particle whose decay exhibits direct CP violation).
The effect will be negligible for $p>\varepsilon f\sim m_\eta$, since for such large
characteristic momenta, the pseudo-Goldstone boson is effectively
massless.~\footnote{We thank Richard Hill for raising this puzzle, and
Ben Grinstein for his physical interpreation.}

In the hypercharge model, there is one exact global ${\rm U}(1)'$
generated by\\ $Y'={\rm diag}(1,1,-4,1,1)/10$, which is spontaneously broken.
A single term is sufficient to lift the Goldstone boson mass, as we have discussed in
the previous section.
However, only in the presence of at least two different terms, a physical CP-odd
phase would arise.
A simple choice would then be the following:
\beq
   \delta\lag_{\rm SCPV} = \varepsilon f^4\lt(a\Sigma_{33}+b\Sigma_{33}^2\rt)+{\rm c.c.},
\eeq
where we take $\varepsilon,a,b$ to be real, and $a,b$ are $\ord(1)$ whereas $\varepsilon$
must be loop suppressed, as we have discussed in the previous section.
This results in the following tree-level potential for $\eta$:
\beq
   V_\eta=2\varepsilon f^4\lt(a\cos\frac{2\eta}{\sqrt{5}f}+b\cos\frac{4\eta}{\sqrt{5}f}\rt)+\ldots
\eeq
This potential is minimized for
\beq
   \vev{\eta}=\frac{\sqrt{5}f}{2}\arccos\lt(\frac{-a}{4b}\rt)\qquad {\rm if}\quad \lt|\frac{a}{4b}\rt|<1,
\eeq
which is of order one if we assume no hierarchy between $a$ and $b$.

Two final comments are in order before we show how the above CP-odd phase
shows up among SM fields.
First, we note that $\eta$ is odd under T-parity, and therefore
a non-zero $\delta$ in the T-parity version of the model
would have to be further suppressed,
as it implies spontaneous breaking of T-parity.
The second comment is about the possibility of CP violation from an overall phase,
$\Sigma_0\to e^{i\alpha}\Sigma_0$.
This phase is related to an overall U$(1)$ which commutes with SU$(5)$.
Therefore, none of the SU$(5)$ Goldstone bosons transform and we conclude that
the overall U$(1)$ is not relevant as a symmetry transformation.
This also means that there is no dynamical field whose VEV is related to that phase.
CP violation from such phases is usually considered explicit, not spontaneous.
Once we include the related Goldstone boson, $\eta'\equiv{\rm Tr}\Pi$, the
overall phase becomes related to spontaneous CP violation.
This amounts to adding a U$(1)$ factor, along with its Goldstone boson to the chiral lagrangian.
We will not consider this issue further, since we find it unrelated to the rest of the discussion.

\subsection{CP Violation in Non-renormalizable Couplings}
\label{sec:CPVSM}

Having found possible modifications of the Littlest Higgs which allow for
spontaneous CP violation, it is worth asking what would be the effect of
CP violation beyond the SM on the SM sector.
Following~\cite{HPR}, we will focus on CP violation in dimension-six couplings of the
SM Higgs to quarks, ignoring other manifestations of CP violation on the SM sector.
Assuming that the low energy effective theory is that of the SM,
new CP violation involving the SM Higgs and quarks would arise predominantly in the
dimension-six operators~\cite{HPR}
\beqa  \label{eq:Zterms}
   \Delta\lag &=& \frac{Z^u_{ij}}{f^2}\bar Q_i\tilde Hu_jH^\dagger H
   +\frac{Z^d_{ij}}{f^2}\bar Q_iHd_jH^\dagger H
   +\frac{Z^{\ell}_{ij}}{f^2}\bar L_iH\ell_jH^\dagger H,\nonumber\\
   &+& \frac{Z^H}{f^2}(D_\mu H)^\dagger D^\mu(H\, H^\dagger H)+{\rm c.c.},
\eeqa
where $f$ is the new physics scale, \ie the spontaneous symmetry breaking
scale of the Little Higgs non-linear sigma model.\footnote{
Note that the last term in Eq.(\ref{eq:Zterms}) can always be shifted away by a non-linear
field redefinition $H\to H\lt(1-\frac{Z^H}{f^2}H^\dagger H\rt)$.
To leading order, such field redefinition mimics replacing $Z^f$ with $Z^f-Z^H$.
Since the authors in~\cite{HPR} assume $Z^H=0$, one has to replace
$Z^f$ by $Z^f-Z^H$ in their results in order to use them correctly.}

The lagrangian (\ref{eq:Zterms}) arises from Eq.(\ref{eq:yukawa}) once we expand $\Sigma$
in terms of the SM Higgs field.
%
%
In this framework, new CP violation (\ie CP violation beyond the CKM phase)
appears as relative phases between the new couplings and the SM Yukawas.
%
%
%
In the Littlest Higgs, the expansion of $\Sigma$ alone in the Yukawa term
does not give rise to relative phases between the coefficients of $H$ and $H H^\dagger H$.
Therefore the only way a phase will show up, would be from two different Yukawa
terms differing in both the expansion coefficients and an overall phase.
However, as we show in appendix~\ref{sec:expansion},
the Littlest Higgs model does not allow for different expansion coefficients,
using any Yukawa term which preserves SU$(3)_1$.
It follows that having a different expansion requires a Yukawa term which
does not respect collective symmetry breaking.
A qualifying Yukawa term is the standard one with a $\Sigma_{33}$ insertion, just
like the one discussed in the previous section.
Such Yukawa term would have to be suppressed in order to keep the SM Higgs light.

We conclude that although we have shown how to get an $\ord(1)$ physical phase
$\delta$ in the Hypercharge model, the phase appearing in Higgs - SM fermions
interactions is suppressed, of order $\varepsilon\delta$.
This is a consequence of the constrained nature of
the SU$(5)$ structure, and is by no means generic.  In order to confirm that, in the next section
we present an SU$(6)/$SO$(6)$ version of the Littlest Higgs.

%

\section{An ${\rm SU}(6)/{\rm SO}(6)$ variant}
\label{sec:su6}

We have found that in ${\rm SU}(5)/{\rm SO}(5)$ Little Higgs models,
spontaneous CP violation requires spoiling collective symmetry breaking,
%
However, this seems to be due to the minimal nature of the SU$(5)/$SO$(5)$ Littlest Higgs.
Once we consider a larger group, it becomes easier to find more
global generators satisfying the two conditions.  We illustrate this
with an ${\rm SU}(6)/{\rm SO}(6)$ version of the Littlest Higgs.
Note that we do not attempt to give a full description of this model and its phenomenology.
Rather, we give a preliminary analysis aimed at the basic features, namely,
a successful mechanism
for suppressing the electroweak scale, lifting all the Goldstone bosons, and
a possible $\ord(1)$ spontaneous CP violation.

We gauge an $[$SU$(2)_1\times$U$(1)]^2$ subgroup of SU$(6)$, generated by
\beqa T_1^i &=&
\pmatrix{\sigma^i/2 & & \cr & {\bf 0}_{2\times 2} & \cr & & {\bf
    0}_{2\times 2}}, \quad
Y_1={\rm diag}(2,2,-1,-1,-1,-1)/6; \nonumber\\
T_2^i &=& \pmatrix{{\bf 0}_{2\times 2} & & \cr & {\bf 0}_{2\times 2} &
  \cr & & -\sigma^{a*}/2}, \quad Y_2={\rm diag}(1,1,1,1,-2,-2)/6.
\eeqa 
This gauging leaves an exact global
${\rm SU}(2)_M$ symmetry which acts on the~(3,4)~block. 
A vacuum which minimizes the effective potential generated by gauge interactions takes the form
\beq 
\label{eq:std-vac-su6}
\Sigma_0=\pmatrix{0 & 0 & \id\cr 0 &
  V & 0\cr \id & 0 & 0}, 
\eeq 
where $V$ may be parametrized as
\beq
   V = \pmatrix{e^{i\alpha}\cos\theta & i\sin\theta \cr
              i\sin\theta & e^{-i\alpha}\cos\theta}=V_{1/2}V_{1/2}^T,\qquad
   V_{1/2}=e^{i\alpha\sigma^3/2}e^{i\theta\sigma^1/2}.
\eeq
This VEV breaks spontaneously the exact global ${\rm SU}(2)_M$ to SO$(2)_M$.
%
The pseudo-Goldstone bosons are parametrized using
\beq
   \Sigma=e^{2i\Pi/f}\Sigma_0,
\eeq
where
\begin{equation}
	\Pi =\pmatrix{& \HH & \phi\cr \HH^\dagger &V_{1/2}\EE V_{1/2}^\dagger &V\HH^T\cr
                 \phi^\dagger &\HH^\ast V^\dagger &},
         \quad \HH = \frac{1}{\sqrt{2}}\lt(H|K\rt),
	\quad \EE =\pmatrix{\sigma &\rho\cr \rho &-\sigma}.
\end{equation}
Note that both $H$ and $K$ carry the quantum numbers of the SM Higgs, whereas $\sigma$ and $\rho$ are SM singlets, and $\phi$ is a complex triplet.
%
The gauge interactions break collectively SU$(4)_1$ and SU$(4)_2$, which protect both doublets
$H$ and $K$ from quadratically divergent mass parameters.
They also leave SU$(2)_M$ unbroken, such that the SM singlets $\rho$ and $\sigma$ remain massless
at this stage.

Note that a quartic coupling $(H^\dagger H)^2$ is not forbidden by collective symmetry breaking,
since the field $\phi$ transforms in such a way that $\Tr\lt|\phi+i/(2f)\, HH^T+\ldots\rt|^2$ remains invaraint.
We compute the quadratic divergent part of the CW potential to verify this in Appendix~\ref{sec:CWPot}. 

In the fermion sector, we introduce the following lagrangian:
\beqa
\label{yuk-su6}
   -\lag_f &=& f\bar \chi_{Li}\left(\lambda_1\Omega_1^i+\lambda_2\Omega_2^i\right) t_R
   +\lambda'f \bar t'_L t'_R,\nn\\
   \Omega_1^i &=& \bar\Sigma^{i4}\Sigma_{44}, \qquad
   \Omega_2^i=\epsilon_{jk\ell}\epsilon_{xy}\bar\Sigma^{ij}\bar\Sigma^{kx}\bar\Sigma^{\ell y},
\eeqa
where $i,j,k,\ell$ run through 1,2,3 and $x,y$ from 5 to 6, and $\chi^i$ includes both the left handed
quark doublet and $t'_L$, as usual - see section~\ref{sec:LLH}.
The Yukawa terms and the mass term break SU$(3)_1$ and SU$(4)_2$ collectively,
such that one doublet, $H$, remains light.
The other doublet, $K$, becomes heavy since it is only protected by SU$(4)_{1,2}$ which are broken
non-collectively by the Yukawa terms.
The Yukawa terms also break SU$(2)_M$ which protects the SM singlet $\rho$,
thus lifting its mass to $\ord(f)$.  Since there are two different spurions which break
this symmetry, we expect spontaneous CP violation from $\theta\sim \ord(1)$.

Note, however, that Eq.(\ref{yuk-su6}) cannot break the SU$(2)_M$ generator $\text{diag}(0,0,1,-1,0,0)$,
since this generator violates the condition from the previous section:
it is an SU$(4)_2$ generator which is also in the span of $\lt\{T^8_{\text{SU}(3)_1},Y,Y'\rt\}$.
This is also manifest in the one-loop effective potential, whose quadratically divergent
term is given by
%
\beq
V_{\text{CW}}(\Sigma)=\frac{\kappa\Lambda^2}{16\pi^2} \Tr MM^\dagger,
\label{eq:CW6}
\end{equation}
where
\beq
M=M(\Sigma)=f\lt(\lambda_1\Omega_1+\lambda_2\Omega_2\rt).
\eeq
Here, the precise value of $\kappa$ depends on unknown physics near the cutoff, and
we have assumed real values for $\lambda_{1,2}$ in order to study the case of
purely spontaneous CP violation.
Using the explicit form of $\Sigma_0$ in Eq.(\ref{eq:std-vac-su6}), this yields
\beq
V_{\text{CW}}=\kappa f^4\cos^2\theta\lt(\lambda_1^2+4\lambda_2^2-\lambda_1^2\cos^2\theta\rt).
\eeq
We can distinguish between two cases:
\ben
   \item
      For $\kappa>0$, the minimum lies at $\theta=\pm\pi/2$.  In that case there is no CP violation,
      since the phase can be removed by the field redefinition
      \beq
         \Sigma\to \exp\lt(\mp T^1_{{\rm SU}(2)_M}\pi/2\rt),
      \eeq
      where $T^1_{{\rm SU}(2)_M}\equiv {\rm diag}\lt(0, \sigma^1/2 , 0\rt)$.
      The potential becomes
      \beqa
         V_{\rm CW}\to && \mp
         \kappa f^4\sin^2\theta\lt[\lambda_1^2
         +4\lambda_2^2\pm\lambda_1^2\sin^2\theta\rt]\nonumber\\
         &&=\mp\kappa f^4\lt(1-\cos^2\theta\rt)\lt[\lambda_1^2
         +4\lambda_2^2\pm\lambda_1^2\lt(1-\cos^2\theta\rt)\rt].
      \eeqa
      Since this field redefinition is not a symmetry, the potential have changed,
      but using the same variable $(\cos^2\theta)$, its coefficients
      remain real valued, while the minimum is now at $\theta=0$,
      hence the field redefinition has removed the phase successfully
      from the lagrangian and it cannot be physical.
   \item
      For $\kappa<0$, the potential is minimized at
      \beqa
         \cos\theta &=&\pm\sqrt{\frac{\lambda_1^2+4\lambda_2^2}{2\lambda_1^2}}\qquad {\rm if}
      \lt|\lambda_1\rt|\ge2\lt|\lambda_2\rt|,\nonumber\\
      \theta &=& 0, \pi\qquad\qquad\qquad {\rm if} \lt|\lambda_1\rt|\le2\lt|\lambda_2\rt|.
      \label{eq:CPVphase-SU6}
      \eeqa
      In the former case, there is a physical CP-odd phase in the vacuum, while in the latter,
      the phase can be removed by a field redefinition.
\een
We conclude that if the UV completion is such that $\kappa<0$, the loop effects are sufficient to generate
a generically large CP-odd phase.   In any case, a phase may be generated also at tree level by introducing
a term of the same form as~Eq.(\ref{eq:CW6}) with a negative coefficient.

As expected, the potential does not stabilize $\alpha$ and this will persist for all the terms in the effective potential,
due to the unbroken U$(1)$ symmetry generated by diag$(0,0,1,-1,0,0)$.
Stabilizing $\alpha$ can be done easily, by introducing a small non-collective breaking term, such as
\beq
   \lag_X=\varepsilon f^4\bar\Sigma^{33}\Sigma_{44} + \text{c.c.}. 
   \label{eq:lagX}
\eeq
Similar to the hypercharge model, we will have to take
$\varepsilon \lsim 1/(4\pi)^2$, which fixes
the mass of $\sigma$ to be around or below the Higgs mass.
%
\begin{figure}
	\centering
	\includegraphics[width=0.42\textwidth]{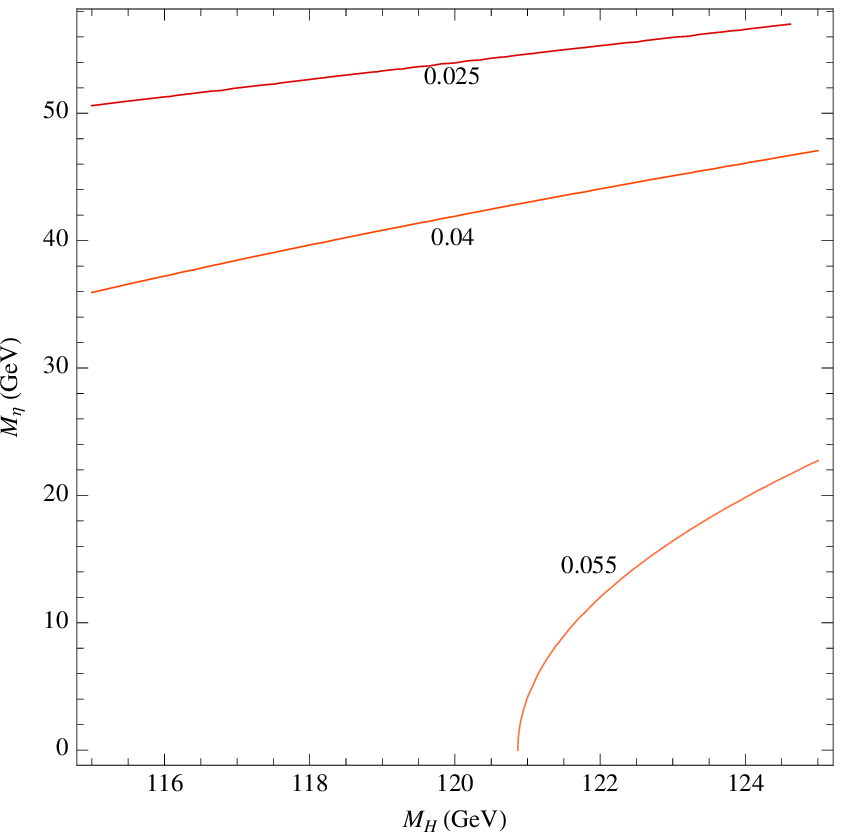}
	\includegraphics[width=0.42\textwidth]{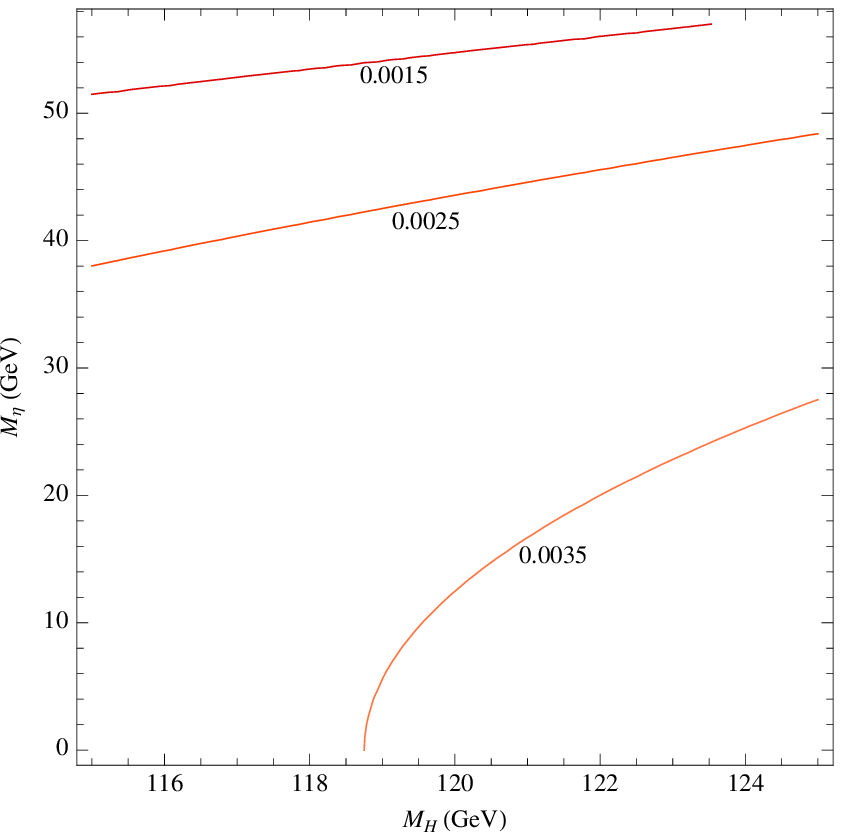}
	\caption{\footnotesize Contours of the branching ratio $BR(h\to\sigma\sigma)$ in the $m_\sigma-m_H$ plane, for $f=1.2$~TeV (left) and $f=2.4$~TeV
	(right).  Again, the dominant mode in this mass range is still $b\bar b$,
	but the new mode $\sigma\sigma$ becomes significant throughout the parameter space.	\label{fig:ratioSU6}}
\end{figure}
%
The light singlet $\sigma$ could alter the Higgs phenomenology,
by playing a role which is similar to the role of the singlet $\eta$ in the
hypercharge model, although the branching ratio for $h\to\sigma\sigma$ at low Higgs mass is slightly lower than the corresponding branching ratio in the hypercharge model (see Fig.~\ref{fig:ratioSU6}).

Unlike the hypercharge model, here an $\ord(1)$ phase would show up in the dimension-six couplings.
Expanding Eq.(\ref{yuk-su6}) in terms of $H$ yields
\beq
   \sqrt{2}e^{-i\alpha}\cos\theta (2i-\sin\theta)\lt[1+\lt(\frac{1}{3}\frac{2i+\sin\theta}{2i-\sin\theta}\rt)
   H^\dagger H\rt]\overline{Q}_{3L} \tilde H t_R.
   \label{eqn:top-yukawa}
\eeq
Note that only $\theta$ is manifested in the SM sector as a relative phase in the
$Z$ couplings, and moreover, the
resulting phase in the low energy lagrangian is not suppressed by the small
parameter $\varepsilon$.
%
%
We conclude that the SU$(6)/$SO$(6)$ model admits spontaneous CP violation from phases in the VEV.
Unlike in the SU$(5)/$SO$(5)$ hypercharge model, the resulting phase between
$\tilde{H}$ and $\tilde{H}H^\dagger H$ can be $\ord(1)$.
%




\section{Conclusions and Further Implications}

In this work we have discussed the tension between lifting Goldstone bosons and 
collective symmetry breaking.
We showed that such tension is present in the SU$(2)\times$SU$(2)\times$U$(1)$
model.  This model has a Goldstone mode which acquires mass only via terms that
spoil collective symmetry breaking.
Such terms must be suppressed in order to keep the SM Higgs boson light,
implying that the related pseudo-Goldstone bosons must be light too.
This may lead to interesting collider phenomenology, such as non-standard Higgs
decays.

Once collective symmetry breaking is spoiled, even by a small parameter, CP
invariance may be broken spontaneously, inducing a large CP-odd phase.
However, due to the minimal nature of the SU$(5)$ structure,
the phase which appears in interactions among SM fields is
suppressed.
We have shown that this difficulty is lifted in an SU$(6)$/SO$(6)$ model, where an
$\mathcal{O}$(1)
phase may give rise to observable effects in non-renormalizabe couplings of the
SM Higgs to quarks.

The $\ord(1)$ spontaneous CP violating phase in Eq.(\ref{eqn:top-yukawa}) would
contribute to electric dipole moments (EDMs)
and will be detectable in the next-generation of electric dipole moments
experiments~\cite{HPR}.

At this level of discussion, we did not suggest experimental ways to distinguish
between spontaneous and explicit CP violation.
In the case of continuous symmetry breaking, there are Goldstone bosons
associated with the continuous set of vacua.
Since CP is a $Z_2$, its spontaneous breaking implies the existence
of two equivalent vacua.~\footnote{
The two vacua cease to be equivalent once explicit
CP violation is added. In our case, this is introduced by the usual
Kobayashi-Maskawa phase.}
Indeed, it is evident from Eq.(\ref{eq:CPVphase-SU6}) that there are
two values for $\theta$ which lead to the same phase in Eq.(\ref{eqn:top-yukawa}).
The doubling of vacua might imply the existence of domain walls in
the universe, once the temperature had dropped below the breaking scale
$f$.  This poses a problem for the cosmology of the model, which
can be avoided if the reheating temperature after inflation is lower than
$f$, or if there is additional explicit CP violation - which would tilt the potential,
making one of the vacua the true vacuum.

A related issue is whether the phase from spontaneous CP violation can
contribute to successful
electroweak baryogenesis.
This depends on other features of the model,
such as the scale $f$ and the sign and size of higher dimension terms in 
the effective potential.
Note that unlike Nelson-Barr models which are renormalizable,
in Little Higgs models the proximity of the UV completion does not permit
predictive statements regarding this issue.
The investigation of the above issues, as well as a detailed analysis
of collider phenomenology, is left for future work.

\acknowledgments
We thank Ben Grinstein for collaboration at the beginning of this 
work, and for insightful discussions.
We are also grateful to Howie Haber, Richard Hill, and Michael Trott
for helpful conversations, and to Tim Tait for reading the manuscript.
This work was supported in part by the US Department of Energy under contract
DOE-FG03-97ER40546.

\appendix
\section{Expansion Coefficients of $\Omega$ in Littlest Higgs and SU$(3)_1$}
\label{sec:expansion}
Here we will show that any two Yukawa terms which are invariant under $SU(3)_1$ have the same expansion coefficients for $H^\dagger$ vs. $H^\dagger H H^\dagger$ (up to an overall constant).
The expression for $\Omega^i$ is given by
\beqa
	\Omega^I &=& \delta^{i3}\left[a_0 + \frac{a_1}{f}\eta +\frac{a_2}{f^2}\eta^2
	+ \frac{a_3}{f^2}H^\dagger H+\frac{a_4}{f^2}\Tr(\phi^\dagger\phi)+\ldots\right]\nn\\
	&\qquad& + \delta^{i\alpha}
	\left[\frac{b_1}{f}H^\dagger + \frac{b_2}{f^2}\eta H^\dagger+\frac{b_3}{f^2}H^\dagger\omega +\frac{b_4}{f^2}h^T\phi^\dagger+\frac{b_5}{f^3}(H^\dagger H)H^\dagger+\ldots\right]^\alpha,
	\label{eq:Omegaa}
\eeqa
where $i = 1,2,3$ and $\alpha=1,2$. Now we will show that the coefficient $b_1$ and $b_5$ are completely determined by $a_0$.
Consider an $SU(3)_1$ transformation generated by
\begin{equation}
	\Lambda =\pmatrix{0 &\lambda &0\cr \lambda^\dagger &0 &0\cr 0 &0 &0\cr}
\end{equation}
The Goldstone bosons transform nonlinearly under $\Lambda$.  Let us define
\beq
	\delta\Pi = \delta\Pi^{(0)}+\delta\Pi^{(1)}+\delta\Pi^{(2)}+\ldots,
\eeq
where
\beqa
	\delta\Pi^{(0)} &=& \frac{f}{2}\left(\Lambda+\Sigma_0\Lambda^T\Sigma_0\right),\nn\\
	\delta\Pi^{(1)} &=& \frac{i}{2}\left(\Lambda\Pi - \Pi\Lambda + \Pi\Sigma_o\Lambda^T\Sigma_0-\Sigma_0\Lambda^T\Sigma_0\Pi\right),\nn\\
	\delta\Pi^{(2)} &=& \frac{1}{6f}\left(-\Pi^2\Lambda+2\Pi\Lambda\Pi-\Lambda\pi^2-\Pi^2\Sigma_0\Lambda^T\Sigma_0+2\Pi\Sigma_0\Lambda^T\Sigma_0\Pi-\Sigma_0\Lambda^T\Sigma_0\Pi^2\right).
\eeqa
In terms of the component fields we have
\beqa
	 \label{eq:fieldtrans}
	\delta H &=&  \frac1{\sqrt{2}}f\lambda + \frac{i}{\sqrt{2}}\left(-\omega\lambda+\phi\lambda^\dagger+\frac{5}{\sqrt{20}}\eta\lambda\right)+ \frac{1}{6\sqrt{2}f}\left[(H^\dagger\lambda+\lambda^\dagger H)H - 2(H^\dagger H)\lambda\right]+\ldots,\nn\\
	\delta\phi &=& \frac{i}{2\sqrt{2}}\left(\lambda H^T + H\lambda^T\right)+\ldots,\qquad
	\delta\eta = -\frac{i\sqrt{10}}{4}\left(H^\dagger\lambda-\lambda^\dagger H\right)+\ldots,\nn\\
	\delta\omega &=& \frac{i}{2\sqrt{2}}\left(\lambda H^\dagger-H\lambda^\dagger\right)-\frac{i}{4\sqrt{2}}\left(H^\dagger\lambda-\lambda^\dagger H\right)+\ldots
\eeqa
Applying $SU(3)$ transformation to $\Omega$ yields
\beqa
	\delta\Omega^i&=& -i\delta^{i\alpha}\left[a_0 + \frac{a_1}{f}\eta +\frac{a_2}{f^2}\eta^2+ \frac{a_3}{f^2}H^\dagger H+\frac{a_4}{f^2}\Tr(\phi^\dagger\phi)\right]\lambda^{\dagger \alpha}\nn\\
	&\qquad&  -i\delta^{i3}\left[\frac{b_1}{f}H^\dagger + \frac{b_2}{f^2}\eta H^\dagger+\frac{b_3}{f^2}H^\dagger\omega +\frac{b_4}{f^2}H^T\phi^\dagger+\frac{b_5}{f^3}(H^\dagger H)H^\dagger\right]\lambda
	\label{eq:LHS}
\eeqa
This must be the same as applying Eq.(\ref{eq:fieldtrans}) to Eq.(\ref{eq:Omegaa})
\beqa
	\delta\mathcal{O}^i &=& \delta^{i\alpha}\frac{b_1}{\sqrt{2}} \lambda^{\dagger \alpha}
	+\frac{1}{f}\left\{\delta^{i3}\left[-\frac{a_1i\sqrt{10}}{4}\left(H^\dagger\lambda-\lambda^\dagger H\right)+\frac{a_3}{\sqrt{2}}\left(H^\dagger\lambda+\lambda^\dagger H\right)\right]\right.\nn\\
	&&+\left.\delta^{i\alpha}\left[-i\frac{b_1}{\sqrt{2}}(-\lambda^\dagger\omega+\frac{5}{\sqrt{20}}\eta\lambda^\dagger + \lambda^T\phi^\dagger)+\frac{b_2}{\sqrt{2}}\eta\lambda^\dagger+\frac{b_3}{\sqrt{2}}\lambda^\dagger\omega+\frac{b_4}{\sqrt{2}}\lambda^T\phi^\dagger\right]^\alpha\right\}  \nn\\
	&& + \frac1{f^2}\delta^{i\alpha}\left\{\frac{b_1}{6\sqrt{2}}
	\left[(H^\dagger\lambda+\lambda^\dagger H)H^\dagger - 2(H^\dagger H)\lambda^\dagger\right]-i\frac{b_2\sqrt{10}}{4}\left(H^\dagger\lambda-\lambda^\dagger H\right)H^\dagger\right.\nn\\
	&&+\left.i\frac{b_3}{4\sqrt{2}}H^\dagger\left(2\lambda H^\dagger-2H\lambda^\dagger-H^\dagger\lambda+\lambda^\dagger H\right)-i\frac{b_4}{2\sqrt{2}}H^T\left(\lambda^\ast H^\dagger + H^\ast\lambda^\dagger\right)\right.\nn\\
	&&+\left.\frac{b_5}{\sqrt{2}}\left[(H^\dagger H)\lambda^\dagger + (\lambda^\dagger H+H^\dagger\lambda)H^\dagger\right]\right\}^\alpha.
	\label{eq:RHS}
\eeqa
Matching the coefficients in Eq.(\ref{eq:LHS}) and Eq.(\ref{eq:RHS}), we get 
\beq
	b_1 = -a_0, \quad b_5 = \frac{i2\sqrt{2}}{3}a_0.
\eeq
Thus any two Yukawa operators $\Omega_1$ and $\Omega_2$ will have the same ratio of the coefficients of  $H^\dagger$ and $H^\dagger HH^\dagger$.

\section{Tree-level Decay Rate for $h\to\eta\eta$}
\label{sec:Gamma}
The kinetic term for the $\Sigma$ field includes the interaction
\begin{equation}
	\frac{1}{f^2}\left[a \eta\partial^\mu\eta\lt(H^\dagger\partial_\mu H +\partial_\mu H^\dagger H\rt)- b\lt(\partial_\mu\eta\rt)^2 H^\dagger H\right]\to \frac{v}{f^2}\left[a\eta\partial^\mu\eta\partial_\mu h-b(\partial_\mu\eta)^2h\right],
\end{equation}
where $v$ is the Higgs vev.  The explicit breaking term, \eg\, Eq.(\ref{eq:sigma33}), includes the interaction
\begin{equation}
	\frac{c M_\eta^2}{f^2}\eta\eta H^\dagger H\to \frac{c v M_\eta^2}{f^2}.
\end{equation}
%
The decay amplitude due to these two terms is
\begin{equation}
	i\mathcal{M}(h\to\eta\eta) = i\frac{v}{f^2}\lt[a p^2+2b(q_1\cdot
	q_2)+cM_\eta^2\rt],
\end{equation}
where $p$ is the momentum of the incoming $h$ and $q_i$ are momenta of the
outgoing $\eta$'s.  In the Higgs rest frame, the amplitude reduces to
\begin{equation}
	i\mathcal{M}(h\to\eta\eta) = i\frac{v}{f^2}M_H^2\lt[a+b+\frac{M_\eta^2}{M_H^2}(c-2b)\rt].
\end{equation}
Thus the rate for $h\to\eta\eta$ is
\begin{equation}
	\Gamma(h\to\eta\eta) = \frac{1}{32\pi}\frac{\sqrt{1-4M_\eta^2/M_H^2}}{M_H}\frac{M_H^4}{f^4}v^2\left[a+b+\frac{M_\eta^2}{M_H^2}(c-2b)\right]^2.
\end{equation}
We include the tree-level decay rate for $h\to b\bar{b}$ for completeness.  The amplitude is 
\begin{equation}
	i\mathcal{M}(h\to b\bar{b}) = 3\frac{ m_b^2}{v^2}\,\Tr\lt(\slashed{p}_1+m_b\rt)\lt(\slashed{p}_2-m_b\rt) = 3\frac{m_b^2}{v^2}M_H^2\left(1-\frac{4m_b^2}{M_H^2}\right).
\end{equation}
Thus the decay rate is
\begin{equation}
	\Gamma(h\to b\bar{b}) = \frac{1}{16\pi}\frac{6m_b^2}{v^2}M_H\left(1-\frac{8m_b^2}{M_H^2}\right)^{3/2}.
\end{equation}

In the hypercharge model $a = b = 5/12$ and $c = 25\sqrt{2}/48$.  For the SU$(6)$/SO$(6)$ model with $\alpha=\theta=0$, we get $a = b = 1/3$ and $c = 17/24$.

\section{1-loop Effective Potential in SU$(6)$/SO$(6)$ Model}
\label{sec:CWPot}
Here we give the one-loop effective potential in the case $\theta = \alpha=0$
and retain only terms relevant for the quartic potential of the Higgs doublet.
The contributions from gauge interactions are
\beqa
	V_{\rm gauge} &=& a (g_1^2+g_1^{\prime2})f^2\,\Tr\left|\phi+\frac{i}{2f}\left(HH^T+KK^T\right)\right|^2\nn\\
	&&+a(g_2^2+g_2^{\prime2})f^2\,\Tr\left|\phi-\frac{i}{2f}\left(HH^T+KK^T\right)\right|^2+\ldots,
\eeqa
where $a$ is an order one constant whose precise value depends on the UV completion.  The contributions form the top quark loop are
\beqa
	V_{\rm top} &=& -\kappa f^4 \left[\lambda_1^2|\Omega_1|^2 + \lambda_2^2|\Omega_2|^2+2\lambda_1\lambda_2{\rm Re}(\Omega_1^\dagger\Omega_2)\right],\nn\\
	|\Omega_1|^2&=& \frac{2}{f^2}\left(2\rho^2+K^\dagger K  \right)\nn\\
	&&-\frac{2i}{f^3}\left(\rho H^\dagger K-\rho K^\dagger H  +K^T\phi^\dagger K -K^\dagger\phi K^\ast\right)+\ord\left(\frac{1}{f^4}\right),\\
	|\Omega_2|^2&=& \frac{8}{f^2}\left(2\Tr\phi\phi^\dagger+ 2\rho^2+K^\dagger K  \right)\nn\\
	&&+\frac{8i}{f^3}\left(\rho H^\dagger K-\rho K^\dagger H  -H^T\phi^\dagger H +H^\dagger\phi H^\ast\right)+\ord\left(\frac{1}{f^4}\right),\\
	{\rm Re}(\Omega_1^\dagger\Omega_2) &=& \ord\left(\frac{1}{f^5}\right),
\eeqa
where $\kappa$ is the order one constant depends on the UV completion.  

\end{document}